\documentclass[
aps,prl,
twocolumn,
reprint,
amsmath,amssymb,
superscriptaddress,
notitlepage
]
{revtex4-2}

\usepackage{graphicx}
\usepackage{amsthm}
\usepackage{cancel}
\usepackage{color}
\usepackage{bbold}
\usepackage{wrapfig}
\usepackage{booktabs}
\usepackage{float}
\usepackage{enumitem}
\usepackage{hyperref}
\usepackage{tikz}
\usepackage{comment}
\usepackage{longtable}
\usepackage{multirow}
\usepackage{hyperref}
\usepackage{cleveref}

\newcommand{\rev}[1]{{\color{black}{#1}}}

\usepackage[normalem]{ulem}

\begin{document}
\global\long\def\pgr{\mathcal{P}_{\text{gr}}}
\global\long\def\pdb{\mathcal{P}_{\text{db}}}
\global\long\def\pov{\mathcal{P}_{\text{ov}}}
\global\long\def\pn{\mathcal{P}_{0}}
\global\long\def\df{d_{\text{f}}}

\title{Phase separation induces congestion waves in electric vehicle charging}

\author{Philip Marszal}
\affiliation{
  Chair for Network Dynamics,
  Center for Advancing Electronics Dresden (cfaed) and Institute for Theoretical Physics, Technical University of Dresden
  01062 Dresden, Germany
}

\author{Marc Timme}
\affiliation{
  Chair for Network Dynamics,
  Center for Advancing Electronics Dresden (cfaed) and Institute for Theoretical Physics, Technical University of Dresden
  01062 Dresden, Germany
}
\affiliation{
  Cluster of Excellence Physics of Life , Technical University of Dresden
  01062 Dresden, Germany
}
\affiliation{
  Lakeside Labs, Lakeside B04b, 9020 Klagenfurt, Austria
}

\author{Malte Schr\"oder}
\affiliation{
  Chair for Network Dynamics,
  Center for Advancing Electronics Dresden (cfaed) and Institute for Theoretical Physics, Technical University of Dresden
  01062 Dresden, Germany
}

\begin{abstract}
Electric vehicles may dominate motorized transport in the next decade, yet the impact of the collective dynamics of electric mobility on long-range traffic flow is still largely unknown. We demonstrate a type of congestion that arises if charging infrastructure is limited or electric vehicle density is high. This congestion emerges solely through indirect interactions at charging infrastructure by queue-avoidance behavior that -- counterintuitively -- induces clustering of occupied charging stations and phase separation of the flow into free and congested stations. The resulting congestion waves \rev{always} propagate forward in the direction of travel, in contrast to \rev{typically} backward-propagating congestion waves known from traditional traffic jams. These results may guide the planning and design of charging infrastructure and decision support applications in the near future.
\end{abstract}

\maketitle

The ongoing transition towards more sustainable mobility centrally relies on electric vehicles to provide low-emission transport. As the number of battery electric vehicles (EVs) grows rapidly
\cite{supp, ieadata, kraftfahrtbundesamt}, EVs may soon become the primary form of individual mobility \cite{ieaevoutlook}.
However, with their limited range and long recharge periods, EVs critically depend on the available charging infrastructure.

Current research on electric mobility thus focuses on cornerstone aspects surrounding the charging process, including the technical implementation of charging and battery technologies \cite{cano2018batteries, schmuch2018performance}, the optimal placement of charging infrastructure \cite{hardinghaus2020real, lam2014electric, xiong2017optimal, wang2018siting, he2019optimal,jochem2019many} and the efficient routing of vehicles within a given infrastructure \cite{qin2011charging, hausler2013stochastic}. In addition, the dependence of electric vehicles on the charging infrastructure has prompted investigations of the interactions, risks, and potential synergies between electric mobility and the larger-scale power grids and energy infrastructure \cite{wolinetz2018simulating, morgenthaler2020three, xu2018planning}. Yet, fundamental aspects of the collective EV charging dynamics
remain poorly understood to date.

In the past, researchers have applied methods of statistical physics, nonlinear dynamics, network science, and complex systems theory to the dynamics of traffic and mobility systems with astounding success. Applications range from describing congestion and phase coexistence in traffic flow and transport processes \cite{richards1956shock, nagel1992cellular, kerner2004three, parmeggiani2003phase, dSouza2005coexisting, loder2019understanding} to understanding the complex interactions in modern networked mobility systems \cite{schadschneider2000statistical, nagatani2002physics, santi2014quantifying, vazifeh2018addressing, karamouzas2014, molkenthin2020, schroder2020anomalous, storch2021incentivedriven}.

In this Letter, we study the collective dynamics of electric vehicles and their interaction with charging infrastructure. We uncover a class of spatio-temporal congestion states for long-distance travel.
In particular, we find congestion waves that are caused solely by indirect interactions of the vehicles with the charging infrastructure in the form of queueing dynamics. We explain the emergence of these waves through phase separation of the charging demand along the available infrastructure. Interestingly, the congestion waves \rev{always} propagate in the direction of travel, not against it as known for \rev{typical \cite{treiber2013traffic, PhysRevE.53.R1297}} congestion waves in traditional traffic flow.

Electric vehicle travel differs from travel by internal combustion engine (ICE) vehicles in two key aspects. Firstly, current EVs possess a typical range of the order of 300~km \cite{evdatabase}, significantly lower than that of ICE vehicles.
Secondly, EVs recover range slowly during charging.
A recharging event at a fast charging station typically lasts more than 20 min \cite{hecht2020representative} and even high-end EVs need about 30 min to recharge 80\% of their battery \cite{teslasupercharger}.
As an example of a characteristic EV recharging rate, we consider $\xi = 480\,\mathrm{km/h}$.
In contrast, internal combustion engine vehicles take mere minutes to refuel, with refuelling rates $\xi_{\mathrm{ICE}} > 10^4 \, \mathrm{km}/\mathrm{h}$. Consequently, refuelling times contribute little to average ICE vehicle travel times while recharging times contribute substantially to EV travel times on long-distance trips.

\begin{figure*}[t!]
    \centering
    \includegraphics{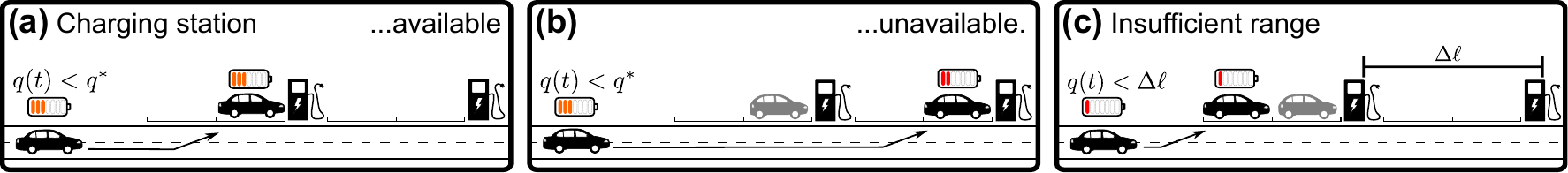}
    \caption{
        \textbf{Charging decisions of EVs.}
        At each charging station, vehicles decide to charge or to continue driving.
        (a) If the remaining range is smaller than a threshold, $q(t)\leq q^*$, the EV charges at an available charging station without a queue.
        (b) If the charging station is currently unavailable and the range is sufficient to reach the next station down the road, the vehicle continues driving to avoid the queue.
        (c) If the vehicles range is insufficient to reach the next charging station, $q(t) < \Delta \ell$, it charges at its current station regardless of queue length.
    }
    \label{fig:schematic}
\end{figure*}

Electric vehicles travelling over long distances $s$ spend characteristic times $t_{\mathrm{driving}}=s/v$ moving and $t_{\mathrm{charging}}=s/\xi$ charging. Their average velocity

\begin{align}
    v_\mathrm{F} = \frac{s}{t_{\mathrm{driving}}+t_{\mathrm{charging}}} = \frac{1}{1/v+1/\xi} \,.
    \label{eq:effective_speed}
\end{align}
is thus only partly determined by the characteristic driving velocity $v$. For instance, for $v = 120\,\mathrm{km/h}$,
the charging time contribution to travel time yields an effective velocity of $v_\mathrm{F}=96\,\mathrm{km/h}$, representing a decrease of $20\%$ due to charging alone. Queues at charging stations exacerbate this effect because waiting times add to the charging times and further reduce the effective velocity. Faced with waiting times comparable in length
to the travel time, EV drivers are likely to employ strategies to avoid queues, similar to queue avoidance behavior observed, for example, during shopping \cite{bennett1998queues} and parking \cite{krapivsky2019simple}.

We consider the basic dynamics of an EV battery gaining and losing charge, specified in terms of the range $q(t)$ available to the vehicle at time $t$. A fully charged vehicle has a maximum range $q_\mathrm{max}$. While driving with a fixed velocity $v$, the available range decreases linearly with time.

At each charging station, the vehicle decides to charge or to continue driving (Fig.~\ref{fig:schematic}). If the available range is below a threshold, $q(t)\leq q^*$, and the charging station is available, the vehicle stops to charge. If there is a queue at the charging station and the vehicle has sufficient range to reach the next station, it continues driving to avoid the queue. If at any time the vehicle does not have sufficient range \rev{$q(t) < \Delta \ell$ to reach the next charging station at distance} $\Delta \ell$, it charges at its current station regardless of the queue length to avoid running out of range in the middle of the road. During the charging process, the vehicle regains range at a constant rate $\xi$ until it is fully charged and continues driving.

To isolate the impact of the charging process on electric vehicle travel, we analyze a basic model where vehicles do not interact or contribute to congestion while driving between charging stations.
We focus on a highway-like system with total length $L$ and periodic boundaries with $K$ equidistant charging stations, each with $m$ charging ports, separated by a distance $\Delta \ell = L/K$.
Vehicles enter the system at a uniformly randomly chosen charging station following a Poisson process with rate $\lambda$ with a uniformly sampled random initial charge $q(0) \in [0,q_\mathrm{max}]$. They travel an average distance $L/2$ and exit the system at another uniformly random charging station. Thus, to complete their journey, a fraction of vehicles will need to recharge at least once.

\begin{figure}[h]
    \centering
    \includegraphics{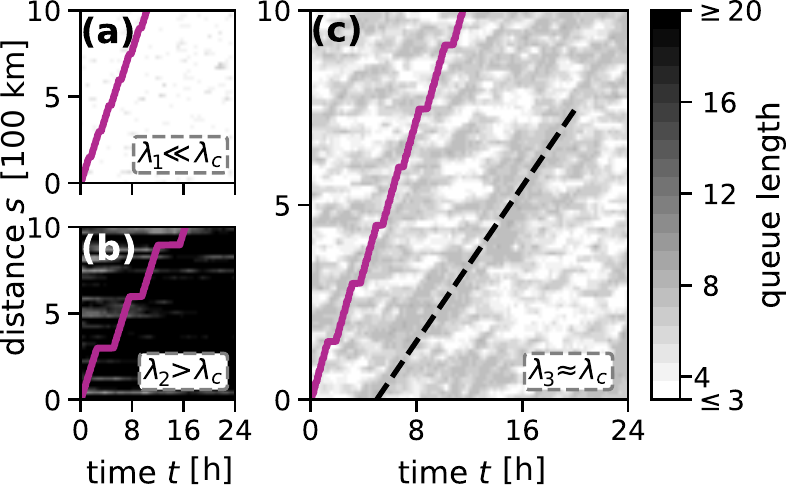}
    \caption{
    {\textbf{Three classes of collective EV charging dynamics.}}
    Space-time plots show the number of vehicles (increasing from white to black) at charging stations distributed along a simulated circular highway with a theoretical capacity $\lambda_\mathrm{c}$ (see main text).
    (a) At low inflow rates (few EVs entering the highway per time, $\lambda_1\ll\lambda_\mathrm{c}$), only short, transient queues appear. The trajectory (purple line) of a single vehicle is characterised by frequent, short charging stops moving with an effective velocity given by Eq.~\eqref{eq:effective_speed}.
    (b) At $\lambda_2 > \lambda_\mathrm{c}$, the system becomes overloaded as vehicles spend a significant amount of time waiting at a charging station; the sample trajectory (purple) exhibits long waiting periods (horizontal) and a significantly smaller effective velocity (average slope).
    (c) A third regime emerges at intermediate inflow rates ($\lambda_3
    \approx \lambda_\mathrm{c} $), exhibiting persistent short queues at some charging stations (grey) that propagate forward in the form of waves (dashed line) at a velocity substantially lower than vehicle velocity (purple). The congested state emerges
    already at $\lambda_3 < \lambda_\mathrm{c}$. Model settings in main text, see \cite{supp} and \cite{modelparams} for additional details.
    }
    \label{fig:phases}
\end{figure}

At a critical rate
\begin{align}
    \lambda_\mathrm{c} = \frac{2mK\xi}{L},
\end{align}
the average consumed range per unit time $\lambda L/2$ exactly matches the maximum total range $mK\xi$ potentially supplied by all charging stations per unit time.
For $\lambda \ll \lambda_c$, the system is in \emph{free flow} with vehicles traveling with an average velocity $v_\mathrm{F}$, unaffected by charging queues (\Cref{fig:phases}a). While short queues may appear randomly even at low numbers of vehicles, they dissolve quickly as vehicles do not enter any particular queue consistently.
In contrast, for $\lambda > \lambda_\mathrm{c}$, the charging infrastructure cannot supply sufficient range for all vehicles. On average, queues grow at every charging station and the system is \textit{overloaded} (\Cref{fig:phases}b).
A third, qualitatively different state emerges at rates just below the critical rate $\lambda_\mathrm{c}$ where self-organized \emph{congestion waves} form across the charging stations (\Cref{fig:phases}c). Like in conventional traffic jams, regions of high vehicle density, here represented by long queue lengths at some charging stations, restrict the flow of vehicles. When queues emerge at localized groups of stations, they propagate along the system with a velocity substantially lower than the effective velocities of the vehicles (slopes of dashed \textit{vs.} purple lines in \Cref{fig:phases}c).

In contrast to conventional traffic jams that \rev{often} propagate backwards, against the direction of travel, charging congestion waves \rev{always} propagate in the direction of travel. Conventional traffic jams grow at the upstream boundary due to the inability of vehicles to pass through each other on the road, forcing cars so stop or slow down when reaching the traffic jam.
In contrast, charging congestion is caused solely by the queue avoidance and charging processes of the vehicles. Vehicles that encounter a group of occupied charging stations continue driving to recharge further downstream due to their reluctance to enqueue. Once the vehicles are forced to enqueue as their remaining charge is too low to reach the next station, they enter queues in the bulk or at the downstream end of the jam. Simultaneously, queues at the upstream end shrink as more vehicles finish charging than enter that queue. Thus, queue lengths grow downstream of the jam and shrink upstream, thereby causing a forward-propagating wave of occupied charging stations. \rev{This mechanism is similar to forward-propagating conventional traffic jams where traffic does not come to a complete standstill in the regions of high vehicle density, moving the congestion forward with it \cite{treiber2013traffic, maerivoet2004non}. Similarly, the electric vehicles in our model still drive in the jammed region though with a lower effective velocity due to charging queues.}
These dynamics remain robust under substantially more general conditions such as heterogeneous properties of both vehicles and charging stations, including varying charging rates, vehicle velocities, maximum ranges, number of ports per station, and station locations (see Supplemental Material \cite{supp}).

We observe the congestion waves more clearly if the total number and thus the overall density of vehicles $\rho$ on the highway is conserved.
To estimate the maximum density $\rho_\mathrm{c}$ of vehicles at which a free-flow state is possible, we consider vehicles as sinks of total charge, i.e.~range, available in the system and charging stations as sources. Jointly, all $K$ charging stations maximally provide new range $mK \xi$ per unit time. Under free-flow conditions, each vehicle consumes charge at a rate matching the effective velocity $v_\mathrm{F}$ defined in \Cref{eq:effective_speed}, in total consuming a range of $L \rho v_\mathrm{F}$. Balancing these source and sink strengths and substituting $K=L/\Delta \ell$ yields the critical density
\begin{align}
    \rho_\mathrm{c} = \frac{m\xi}{v_\mathrm{F}\Delta \ell},
\end{align}
of vehicles on the highway.

\begin{figure}[h]
    \centering
    \includegraphics{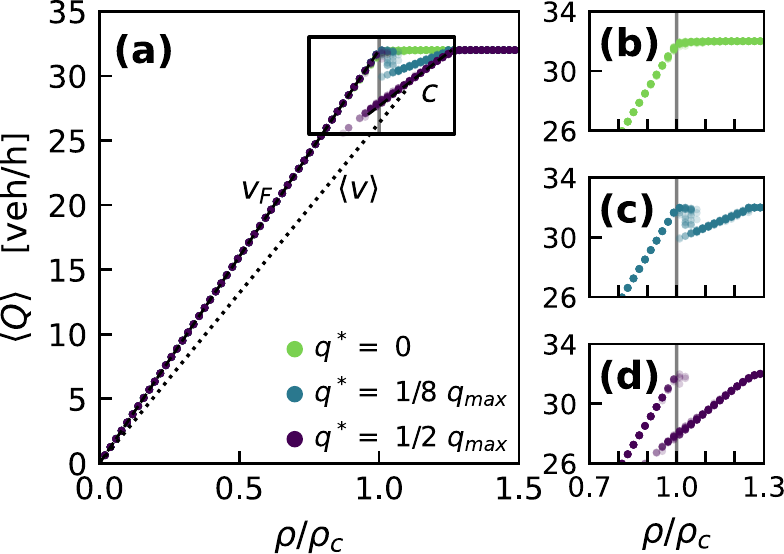}
    \caption{
    \textbf{Congestion waves near the critical density.}
    (a) The fundamental diagram quantifies the flow states as a function of the vehicle density qualitatively observed in \Cref{fig:phases}.
    At low densities the system settles in a free-flow state where $\langle Q\rangle = v_\mathrm{F} \rho$, with an effective velocity $v_\mathrm{F}$ defined by \Cref{eq:effective_speed}. Substantially above the critical density, the system overloads and the flow saturates at a maximum flow $\langle Q\rangle = Q_\mathrm{max}$ set by the total charging rate of the system. At intermediate densities $\rho/\rho_\mathrm{c} \approx 1$, a congested state emerges, reducing the average velocity of individual vehicles to $\langle v \rangle = \langle Q\rangle / \rho$.
    Congestion waves  move at a group velocity $c = \partial\,\langle Q\rangle/\partial\,\rho > 0$ indicated by the local slope.
    (b) The congested state does not emerge if vehicles deplete their batteries completely before charging ($q^* = 0$, light green).
    (c,d) Congestion occurs at lower densities for higher charging thresholds (c, $q^*=1/8~q_\mathrm{max}$, blue and d, $q^*=1/2~q_\mathrm{max}$, purple).
    The fundamental diagram shows the time-averaged flow from 25 realizations for each density in a system with a fixed number of vehicles $N=\rho\,L$ and $m=1$ charging port per station; all other parameters are identical to Fig.~\ref{fig:phases}.
    }
    \label{fig:fundamental}
\end{figure}

The fundamental diagram  (Fig.~\ref{fig:fundamental}) that links the flow $Q(\rho)=v(\rho)\rho$ to the vehicle density, offers two helpful relations for quantifying the properties of traffic flow \cite{treiber2013traffic}. First, it provides the time-average velocity
\begin{align}
    \langle v \rangle = \frac{\langle Q(\rho) \rangle }{\rho}.
\end{align}
of the vehicles. Second, it offers insights about the propagation of density variations of the form $\Tilde{\rho}(x-ct)$ through the group velocity \cite{treiber2013traffic}
\begin{align}
    c = \frac{\partial \langle Q(\rho) \rangle}{\partial \rho}.\label{eq:jamspeed}
\end{align}

At densities $\rho < \rho_\mathrm{c}\, $, queues do not form at all. In this free-flow state, the average velocity is independent of the vehicle density and the flow increases linearly with the number of vehicles in the system. Without queue avoiding behaviour ($q^* = 0$, \Cref{fig:fundamental}b), the system overloads as the density exceeds the critical density $\rho_\mathrm{c}$, where the flow saturates and becomes density-independent, $\langle Q(\rho) \rangle = \xi/\Delta \ell$, because it is limited by the total available charging rate.
Furthermore, since the group velocity $c = 0$, density fluctuations do not propagate in space and the vehicle densities at all charging station are approximately constant in time (compare Fig.~\Cref{fig:phases}b).

With active queue avoidance, $q^* > 0$ (\Cref{fig:fundamental}c,d), the flow dynamic changes fundamentally. Close to the critical density $\rho_\mathrm{c}$, the fundamental diagram exhibits a discontinuity and propagating congestion waves emerge, decreasing the flow regardless of whether the system would be overloaded or in free flow. The discontinuity becomes stronger with larger charging threshold and the flow exhibits bistability: congestion waves may already emerge at densities below $\rho_\mathrm{c}$, effectively decreasing the critical density
before the free-flow state breaks down.

\begin{figure}[h]
    \centering
    \includegraphics{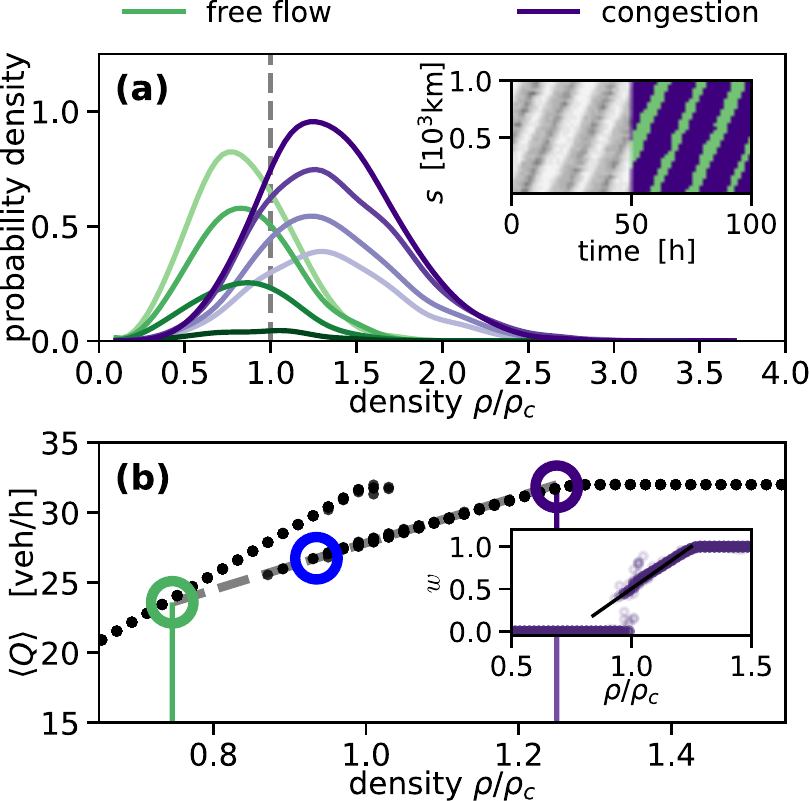}
    \caption{
    \textbf{Phase separation of the flow into a free-flow phase and an overloaded phase.}
    (a) Continuous interpolation of the conditional probability distributions of the vehicle densities in both states. The system separates into a free-flow state with low density (green) and an overloaded state with high density (purple).
    A larger overall vehicle density (darker colored lines) leads to a larger number of charging stations that are simultaneously congested captured in the relative weight of the probability density distributions. The average conditional densities of the respective free-flow and congested states, however, are independent of the overall density (see \cite{supp}). (a, inset) Classification into free-flow and congested states (see \cite{supp} for details).
    (b) Illustration of the phase separation along the tangent in the fundamental diagram. The flow of a system exhibiting propagating congestion at overall vehicle density $\rho_0$ is the weighted average of the flow in the congested phase (purple circle) and the flow in the free-flow phase (green circle).
    (b, inset) Measuring the conditional vehicle densities $\rho_\mathrm{free}$ in free-flow regions and $\rho_\mathrm{jam}$ in congested regions predicts the fraction $w$ of congested stations using \Cref{eq:density_relation}.
    The fraction of congested stations observed (points) matches the prediction (dashed line). Parameters are identical to Fig.~\ref{fig:fundamental} with charging threshold $q_\mathrm{c} = 1/2~q_\mathrm{max}$.
    }
    \label{fig:phase_separation}
\end{figure}

To further understand the spatio-temporal structure of the congestion waves, we classify each station as either congested or free at a given time, taking into account the immediate temporal and spatial neighborhood using a Gaussian mixture model (Fig.~\ref{fig:phase_separation}a, see
\cite{supp} for details). The probability density for finding a certain vehicle density at a station differs between the congested and the free-flow state. The probability densities exhibit peaks below the critical vehicle density $\rho_\mathrm{c}$ in the free-flow state and above $\rho_\mathrm{c}$ in the congested state.

These findings indicate the emergence of phase separation in the system. In a system exhibiting congestion waves at some density $\rho_0$ close to $\rho_c$, the flow splits into two distinct phases of flow, represented by two points on the fundamental diagram. As observed in \Cref{fig:phase_separation}a, the free-flow phase has an average conditional vehicle density $\rho_\mathrm{free} < \rho_c$, the congested flow phase has an average conditional density $\rho_{\mathrm{jam}} > \rho_c$, both insensitive to the total vehicle density $\rho_0$ (see \cite{supp}). These flow states are marked by the intersection of the tangent of the fundamental diagram at the point $(\rho_0, \langle Q(\rho_0) \rangle )$ with the free-flow and overloaded branch, respectively (\Cref{fig:phase_separation}b). Since the conditional densities in the free-flow and congested phase are constant,
the fraction $w$ of congested stations is directly related to the overall vehicle density
\begin{align}
    \rho_0 = w \rho_\mathrm{jam}+(1-w)\rho_\mathrm{free} \, \label{eq:density_relation}
\end{align}
that is the weighted average of the free-flow and congested densities. \Cref{eq:density_relation} predicts the fraction of congested stations $w$ just from the measured densities $\rho_0$, $\rho_\mathrm{free}$ and $\rho_\mathrm{jam}$. Comparing the result to the measured fraction of congested stations confirms this prediction and the phase separation hypothesis (see \Cref{fig:phase_separation}b inset).

In conclusion, our analysis demonstrates that charging dynamics of EVs together with individual queue-avoidance behavior jointly induce spatio-temporal congestion.
\rev{Specifically, we find congestion waves of occupied charging stations without direct interactions of the vehicles on the road. These congestion waves always propagate forward, along the direction travel as vehicles continue to drive past occupied charging stations and effectively enqueue at the downstream end of the jam.
The phase separation into free-flow and overloaded stations underlying these congestion waves is similar to mechanisms of congestion phenomena in conventional traffic jams \cite{treiber2013traffic, pottmeier2002localized} and} other transport processes \cite{parmeggiani2003phase, dSouza2005coexisting} and to the phase separation known from standard thermodynamics such as the liquid-gas transition in van-der-Waals fluids \cite{sethna2021statistical}, where the overall density is a superposition of the densities of one gas and one liquid phase.
If charging infrastructure is limited or EV density is high, such congested states may further increase the overall travel time of EVs during long distance trips, in addition to the already substantial charging times. Charging dynamics of EVs thus constitute an additional dimension
in traffic flow analysis in addition to their impact on power systems \cite{wolinetz2018simulating, morgenthaler2020three, xu2018planning}, becoming increasingly relevant as society transitions towards electric transport.

The observed congestion states emerge exactly around the mean field critical point, reflecting the economic equilibrium where charging stations are maximally utilized but the system is not yet overloaded. While high charging thresholds and vehicle densities strongly promote congestion, the congested states appear for a wide range of (heterogeneous) system parameters and conditions (see \cite{supp} for details). Understanding the dynamics of congestion waves in EV charging infrastructures in further detail for specific infrastructure scenarios may support civil engineers and policy makers to anticipate inefficiencies and implement countermeasures. Our results may already provide conceptual guidance on how to mitigate the problem of congestion waves in EV charging by highlighting the required information to design smart routing and charging suggestions \cite{qin2011charging, hausler2013stochastic}. Moreover, the insights illustrated above highlight one key ingredient to minimize the impact of charging congestion that goes beyond infrastructure development or technological progress: Decreasing the charging threshold, \rev{for example} by strengthening confidence in the vehicle-predicted remaining range or by automating charging decisions, limits the congestion to a small range of vehicle densities.

More generally, our results indicate that emerging technologies not only change individual technical components underlying transport, such as vehicles or infrastructure. Given new forms of interactions among system elements, we also expect a range of unprecedented collective dynamics \cite{holovatch2017complex, RevModPhys.81.591, schroder2020anomalous} emerging in future-compliant mobility systems we need to explore.\\

We thank Christopher Hecht for helpful discussions. M.T. acknowledges support from the German Research Foundation (Deutsche Forschungsgemeinschaft, DFG) through the Center for Advancing Electronics Dresden (cfaed).

\bibliography{main_bib}

\clearpage

\setcounter{figure}{0}
\renewcommand{\thefigure}{S\arabic{figure}}
\renewcommand{\thetable}{S\arabic{table}}
\renewcommand{\theequation}{S\arabic{equation}}

\onecolumngrid
\section{Supplemental Material}





\subsection{Current trends in EV infrastructure}

\subsubsection*{Exponential growth in electric vehicle numbers vs. linear growth in fast charger numbers}
\begin{figure}[h]
    \centering
    \includegraphics{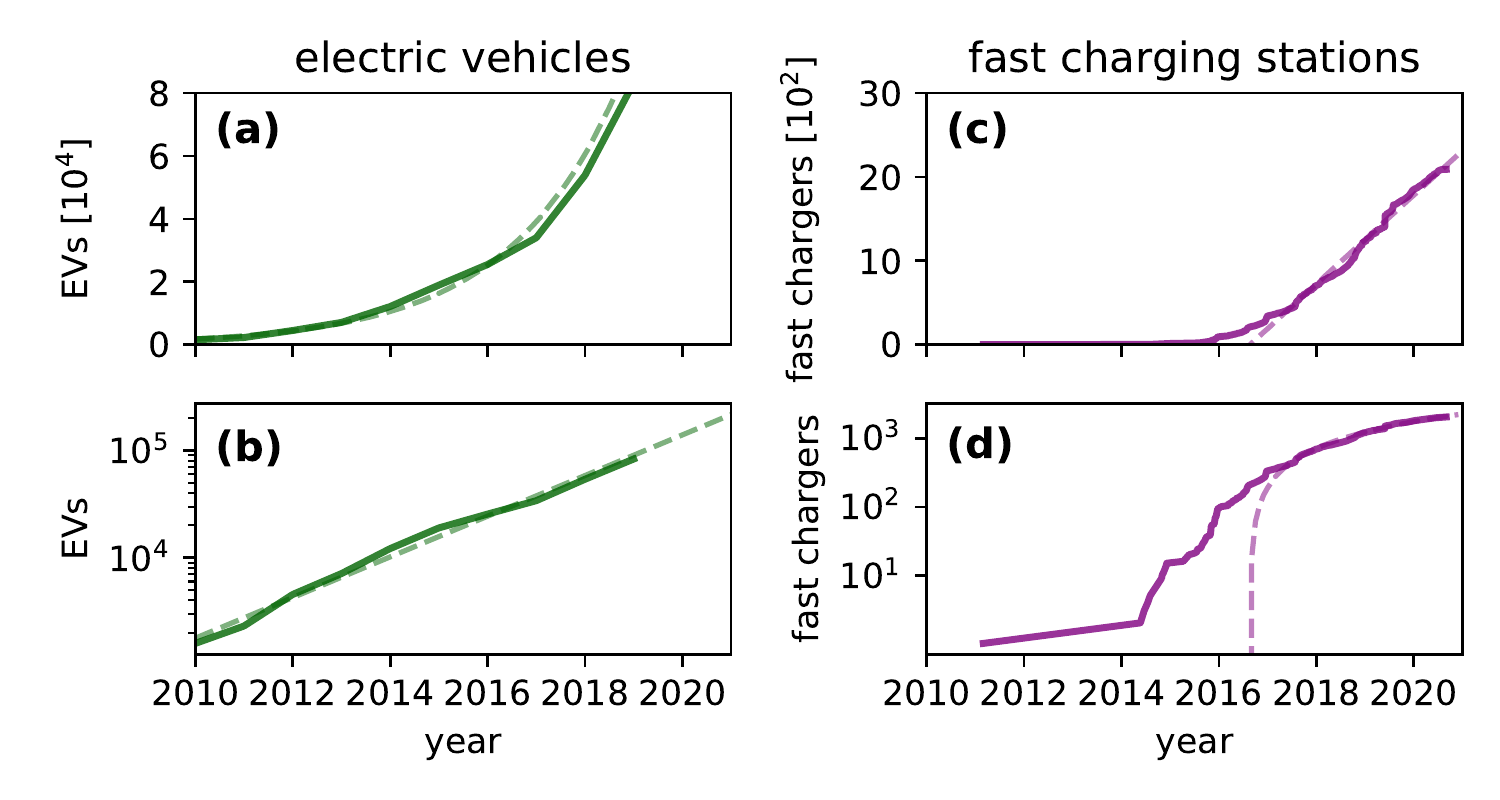}
    \caption{\textbf{Current trends in German electric mobility.} (a,b) The number of electric vehicles is increasing exponentially with almost constant growth rate over the last decade (panel a: linear scale, panel b: logarithmic scale).  An exponential fit (dashed line) yields a growth rate of $0.437,\mathrm{y}^{-1}$ corresponding to a doubling every $1.6$ years. (c,d) Fast charging stations, on the other hand, display only linear growth (panel c: linear scale, panel d: logarithmic scale) over the last four years. This discrepancy in the growth dynamics may lead to a shortage of charging stations and contribute to congestion phenomena emerging in charging infrastructure.\\
    }
    \label{fig:supp:numbers}
\end{figure}

The number of registered battery electric vehicles in Germany has increased steadily over the last decade \cite{kraftfahrtbundesamt}. Data on the number of registered vehicles suggests exponential growth with doubling expected to occur every $1.6$ years (see \Cref{fig:supp:numbers}a,b). As discussed in the main manuscript, sufficiently developed fast-charging infrastructure is required for reliable long distance travel. However, the number of fast chargers in Germany is only increasing linearly \cite{bundesnetzagentur}. Initial rapid increases in charging infrastructure may be attributed to market dynamics and the race of providers to dominate the market. In recent years growth has significantly slowed down to a continuous linear increase in the last three years (see \Cref{fig:supp:numbers}c,d). This growth discrepancy results in an increasing vehicle to charging station ratio, increasing the likelihood the observed congestion dynamics may occur in the future.

\vspace{-0.75cm}

\subsubsection*{Vehicle counts on German motorways}
In Fig. 2 in the Letter we showed the flow states of a simulated highway for different rates of cars. The critical rate of electric vehicles entering the highway with our model parameters is $\lambda_\mathrm{crit} = 256 \mathrm{vehicles/h}$.
Available measurements suggest that the number of vehicles using the A7 per day ranges from 8000 vehicles to 48000 vehicles per day in each direction, depending on the segment of the highway \cite{bast}. Distributed over 16 hours (excluding night time with little traffic) this results in a rate $\lambda$ between 500 and 3000 vehicles per hours. If this traffic would be completely composed of long distance EV journeys, the highway would be well in the overloaded state given the current charging infrastructure.

However in January 2020 only about 0.3\% of vehicles in Germany were battery electric vehicles \cite{kraftfahrtbundesamt}. This translates to an EV rate $\lambda$ between 1.5 and 9 vehicles per hour, a value currently far below the capacity of the highway. As seen above however EV market penetration is certain to increase and hence in the future congestion at charging stations may become more prevalent, especially when the infrastructure continues to grow more slowly than EV usage.

\clearpage

\subsection*{Simulation}
We study the electric vehicle traffic problem described in the main manuscript using event-based simulations. The simulated road consists of a sequence of charging stations connected by road segments. We neglect the interaction between vehicles on road segments, which allows us to view the electric vehicle traffic flow as a sequence of discrete events:
\begin{itemize}
    \item EV leaves a station
    \item EV arrives at a station
    \item EV enters a queue
    \item EV initiates charging
    \item EV finishes charging
\end{itemize}
Each event corresponds to a change of the microscopic system state. Each microscopic state has a duration, that can be analytically calculated. For example:
A vehicle $A$ arrives at charging station $1$ at time $t_0$. The its driver decides not to charge, therefore the next event scheduled will be the vehicle arriving at charging station 2 at time $t_1 = t_0 + d(1,2)\cdot v$. Upon arrival the state variables of the car are updated, i.e. the current range is reduced by the distance driven $q(t_0) \rightarrow q(t_1)$. If charges, then the next event queued is the driver leaving the station at the time $t_2 = t_1 + (q_\mathrm{max}-q(t_1))\cdot \xi$.

With multiple vehicles, a vehicle may have to queue before charging. We implemented the simulation code in python and use the \textit{SimPy} framework for event-based simulation \cite{simpy}, modeling charging stations as multi-server queues, where the number of servers corresponds to the number of charging ports. Working examples of the code used will be made available upon publication.

\subsubsection*{Parameters}
We chose system parameters that reflect current conditions on German motorways, as obtained from OpenstreetMaps \cite{openstreetmap}, osmnx \cite{boeing2017osmnx} and Bundesnetzagentur \cite{bundesnetzagentur}. The average driving velocity is $120\,\mathrm{km/h}$ roughly based on average travel speed on the A7. The distance between subsequent charging stations $l = 15\,\text{km}$ also approximately matches the average distance between charging stations on the German Bundesautobahn A7 ($16.01\,\mathrm{km}$). The number of ports for Fig. 2 (4 ports per station) also reflects the average number of ports at fast charging stations on the Bundesautobahn A7 (4.55 ports per station on average with a minimum of 2 and a maximum of 16).

Charging speed is in general inhomogeneous across vehicles and stations. In \cite{mckinseycharging} the average recharge rate is estimated to be $350\,\mathrm{km/h}$. However EV Manufacturers commonly claim greater charging speeds. Using the claim made by Tesla, that a battery can be charged from 0\% to 80\% in 30 minutes \cite{teslasupercharger}, we estimate a larger average recharge rate: For a typical range of $300\,\mathrm{km}$ (smaller and cheaper cars have less range \cite{evdatabase}, high-end vehicles may have up to $650\mathrm{km}$ range \cite{teslarange}), this results in a charging speed of $480\,\mathrm{km/h}$, or four times the average vehicle speed. We use this higher estimate given that charging speeds are likely to increase in the future. Smaller charging speeds only increase the observed effects as the critical density of vehicles per charging station becomes smaller and individual charging events last longer. Tables \Cref{tab:fig2params,tab:fig3params} summarize the parameters used for the simulations in Fig. 2 and Fig. 3 respectively.

\begin{table}[h]
    \centering
    \caption{Parameters for the simulations in Fig. 2.}
    \label{tab:fig2params}
    \begin{tabular}{l|c|c}
        description & symbol & value \\\hline
         vehicle velocity& $v$ & $120\,\mathrm{km}/\mathrm{h}$\\
         charging rate& $\xi$ & $480\,\mathrm{km}/\mathrm{h}$ \\
         vehicle range& $q_\mathrm{max}$& $300\,\mathrm{km}$\\
         charging threshold& $q^*$ & $150\,\mathrm{km}$ \\
         charging station distance& $\Delta \ell$& $15\,\mathrm{km}$\\
         charging ports per station & $k$ & $4$\\
         highway length & $L$ & $1515\,\mathrm{km}$\\
         vehicle rate & $\lambda_1\,,\lambda_2\,,\lambda_3\,$ & $ 150\, \mathrm{veh}/\mathrm{h}\,\,, 300\, \mathrm{veh}/\mathrm{h}\,\,, 230\, \mathrm{veh}/\mathrm{h}\,\,$ \\
    \end{tabular}
\end{table}

\begin{table}[h]
    \centering
    \caption{Parameters for the simulations in Fig. 3 and Fig. 4.}
    \label{tab:fig3params}
    \begin{tabular}{l|c|c}
         description & symbol & value \\\hline
         vehicle velocity& $v$ & $120\,\mathrm{km}/\mathrm{h}$\\
         charging rate& $\xi$ & $480\,\mathrm{km}/\mathrm{h}$ \\
         vehicle range& $q_\mathrm{max}$& $300\,\mathrm{km}$\\
         charging station distance& $\Delta \ell$& $15\,\mathrm{km}$\\
         charging ports per station & $k$ & 1\\
         highway length & $L$ & $1515\,\mathrm{km}$\\
    \end{tabular}
\end{table}

The emergence of the congestion waves is an inherent effect of the queueing and decision making of drivers and largely independent of the specific system parameters. Our observations thus remain robust over a broad range of parameters as well as for heterogeneous vehicles or infrastructure settings (see below).

\clearpage
\subsection*{Classifying congestion}

\begin{figure}[h]
    \centering
    \includegraphics[width=\textwidth]{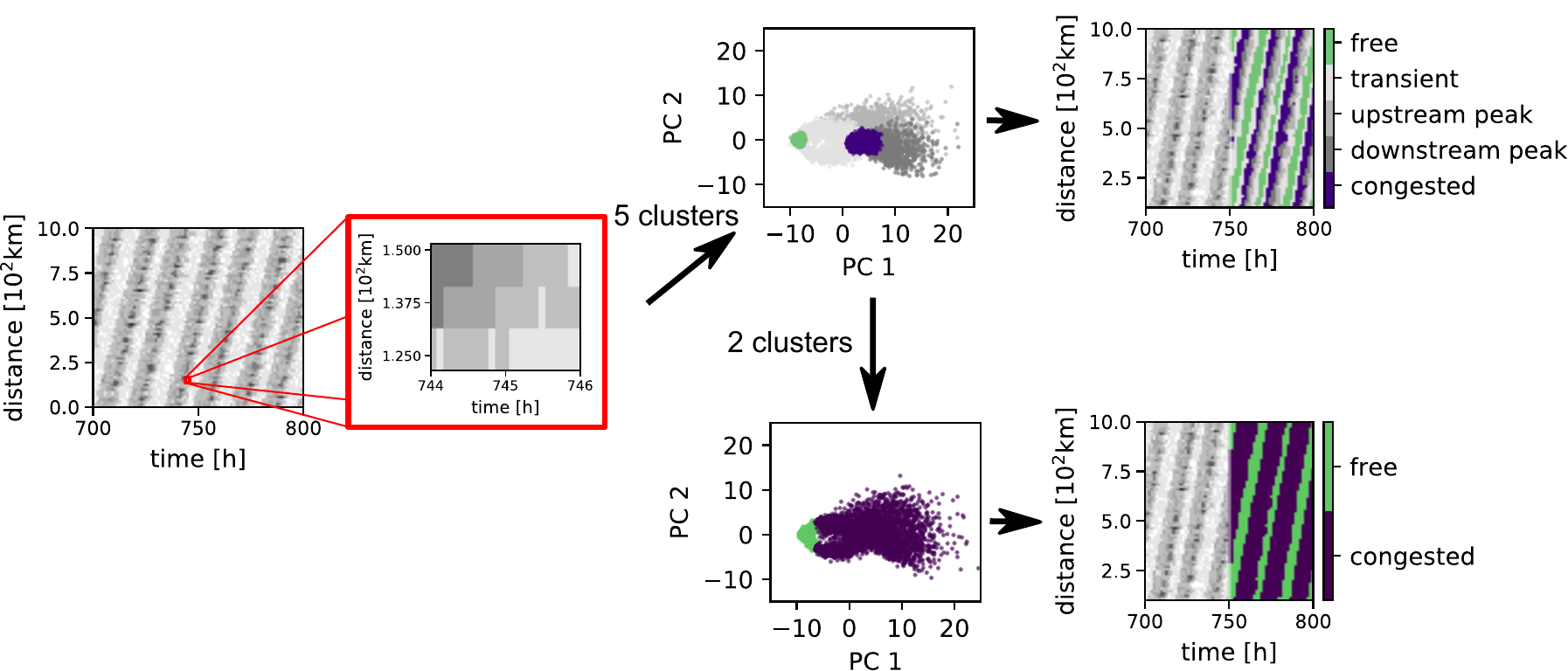}
    \caption{\textbf{Classification of charging station state into congested and free-flow states.} Simulations provide time series for the queue lengths of each charging station with a sampling interval of 6 minutes. The time series is transformed into a list of windows containing queue lengths of a charging station and its immediate neighbouring stations over a time span of 2 hours (to capture potential changes of the queue length given an average charging time of 30 minutes). These windows are interpreted as 63-dimensional vectors. Using principal component analysis, we identify the four main components of the queue lengths, which account for 80\% of the variance in the data set. We fit a Gaussian mixture model consisting of five distributions to the vectorized data, obtaining a consistent division of possible flow states into congested, free flow, transient, upstream peak front and downstream peak front. The latter two describe the peak at the upstream front of the traffic jam. Since transient and peak states only occur in conjunction with congested flow, we restrict our analysis to two states and assign each data point to either the free-flow cluster or the congested cluster, based on likelihood. This results in the two state separation further analyzed in the main manuscript.\\
    }
    \label{fig:supp:classification}
\end{figure}

In order to study the properties of the congestion waves we classify charging stations into congested and free-flow phases using a Gaussian mixture model.
For any station at position $x_0$ at a given time $t_0$, we consider a space-time window $x \in \left\{x_0-1, x_0, x_0+1\right\}$ of immediately adjacent stations and $t \in [t_0-\Delta t,t_0+\Delta t]$ (Fig.~\ref{fig:supp:classification} left). Treating the queue lengths of stations in this window as a vector we fit a 5-component Gaussian mixture model to the vectors obtained from a simulation that exhibits congestion waves. The five components describe the different parts of the congestion waves: the two slopes of peaks occurring at the upstream end of the congestion (darker spots at the bottom of the congested region), a large congested area with constant queue length, free-flow areas with transient, stochastic queues and a not uniquely identifiable transient state (Fig.~\ref{fig:supp:classification} top).

We then choose two of these components to classify the system into only congested and free-flow states. In practice this means, that components corresponding to the boundaries of the congestion and transient states are merged into a single congested state (Fig.~\ref{fig:supp:classification} bottom).

Training one mixture model translates well to other simulations with the same parameters at different average densities due to the fact that the density in the congested state remains constant. This classification is the basis for the investigation of conditional densities and phase separation discussed in the main manuscript (Fig.~4 in the main manuscript). The conditional densities are measured over the integer number of vehicles at a single charging station classified as free flow or congested, respectively. The distribution has been interpolated in Fig.~4 in the main manuscript for easier readability.

\clearpage
\subsection{Probability of congestion}
In Figure 3 of the main manuscript we observe bistability of congestion and free-flow states at the same vehicle density depending on the initial conditions. We quantify the probability for congestion to emerge by sampling the space of initial conditions by initializing cars at uniformly random charging stations along the highway with a uniformly random charge $q\in[0,q_\mathrm{max}]$ and observe which fraction of initial conditions lead to congestion.

\begin{figure}[!h]
    \centering
    \includegraphics[width=\textwidth]{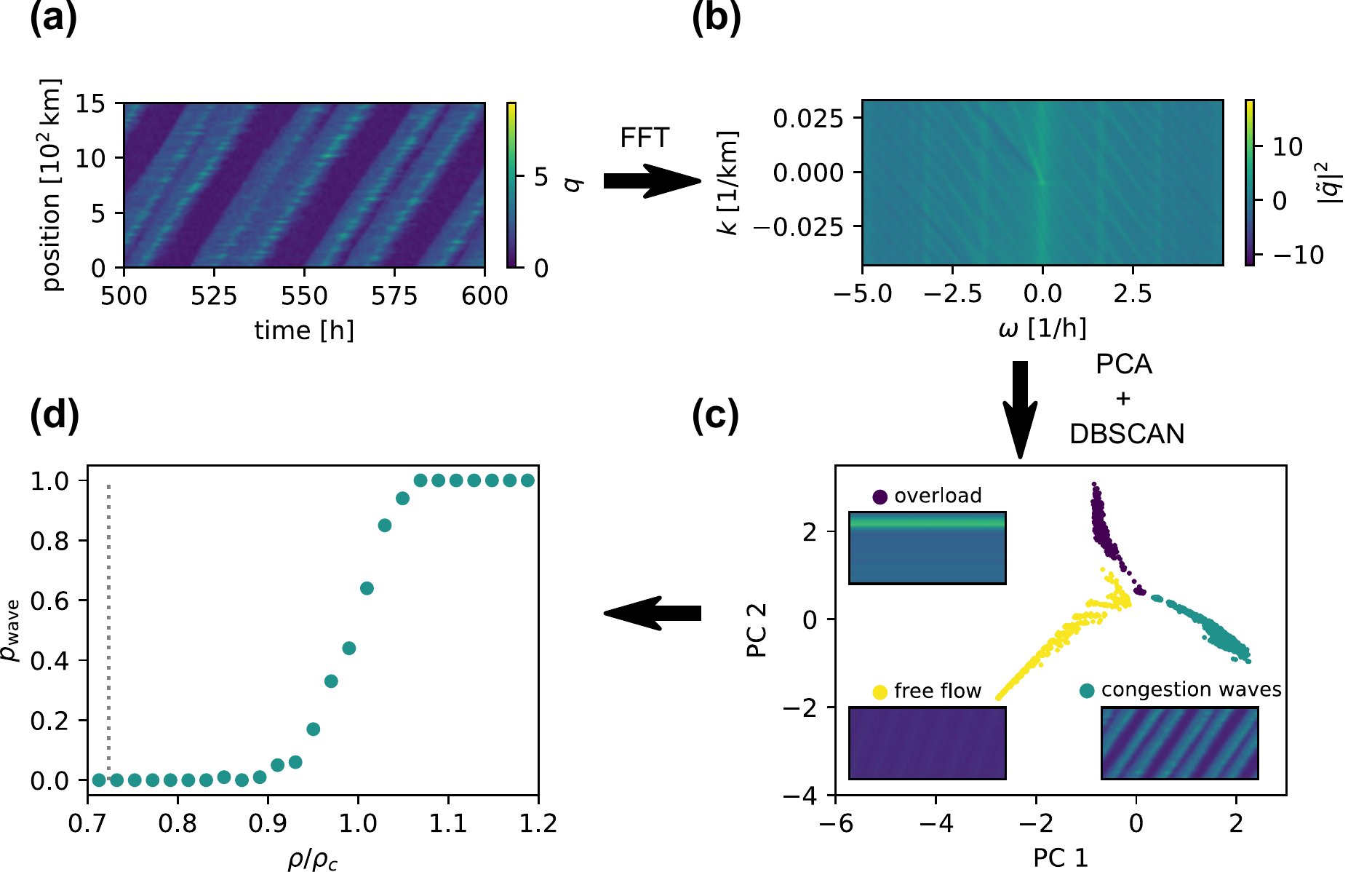}
    \caption{\textbf{Probability of congestion.} (a) Simulations at uniformly randomly sampled initial conditions provide time series data for each charging station. (b) The congestion waves produces clear patterns in frequency-space. (c) The power spectra for all realizations are decomposed into two principal components. Realizations clearly separate into three clusters, which are detected using the DBSCAN method. The three clusters coincide with the three possible states of the System, free flow, congestion waves, and overload. (d) The results of the cluster analysis are used to determine the fraction of realizations leading to congestion waves. The system becomes fully overloaded only for large densities, $\rho/\rho_c > 1.25$.
    }
    \label{fig:congestion_probability}
\end{figure}

We determine whether a system exhibits congestion, by making use of the wavelike nature of states exhibiting propagating congestion [Fig.~\ref{fig:congestion_probability}(a)]. First we perform a two dimensional Fourier transform of the time series data for every realization. We then use the power spectral density as a function of the wave number $k$ and the frequency $\omega$ as the basis for the classification [Fig.~\ref{fig:congestion_probability}(b)]. After performing a principal component decomposition, we identify all realizations to fall onto a pinwheel shape. The three branches of the pinwheel coincide with the three states of the system, free flow, congestion waves and overload [Fig.~\ref{fig:congestion_probability}(c)].
We use a density based clustering algorithm, DBSCAN, to classify the realizations into three clusters. The probability of congestion is then taken to be the fraction of realizations falling within the congestion wave cluster [Fig.~\ref{fig:congestion_probability}(d)].

The probability of congestion strongly increases when the density increases. At densities $\rho/\rho_c>1$ the system is still not certain to exhibit congestion waves. Therefore coexistence exists not only between free flow and congestion waves, but also between overload and congestion waves.

\clearpage
\subsection*{Conditional vehicle densities}
The average densities $\rho_\mathrm{free}$ and $\rho_\mathrm{jam}$ in the free-flow and congested states remain constant when the total density of cars is varied. To show this we identify the simulations which exhibit congestion waves by utilizing the clustering model described in \Cref{fig:congestion_probability}.

We detect whether a simulation exhibits congestion waves by transforming the queue length time series in the manner described in \Cref{fig:congestion_probability}. We then determine the overall state of the system for this simulation through a nearest neighbour search. From the already clustered dataset we find the ten simulations that are closest to the simulation in question. We assume there are congestion waves in this simulation, if a majority of the neighbours are also categorized as exhibiting congestion waves.

For every simulation where congestion waves were detected, we perform a classification approach based on the model developed in \Cref{fig:supp:classification}. We detect five different states, congested, free flow, and three transient state on the edges of the congestion waves, as described above.

\begin{figure}[!h]
    \centering
    \includegraphics[width=0.7\textwidth]{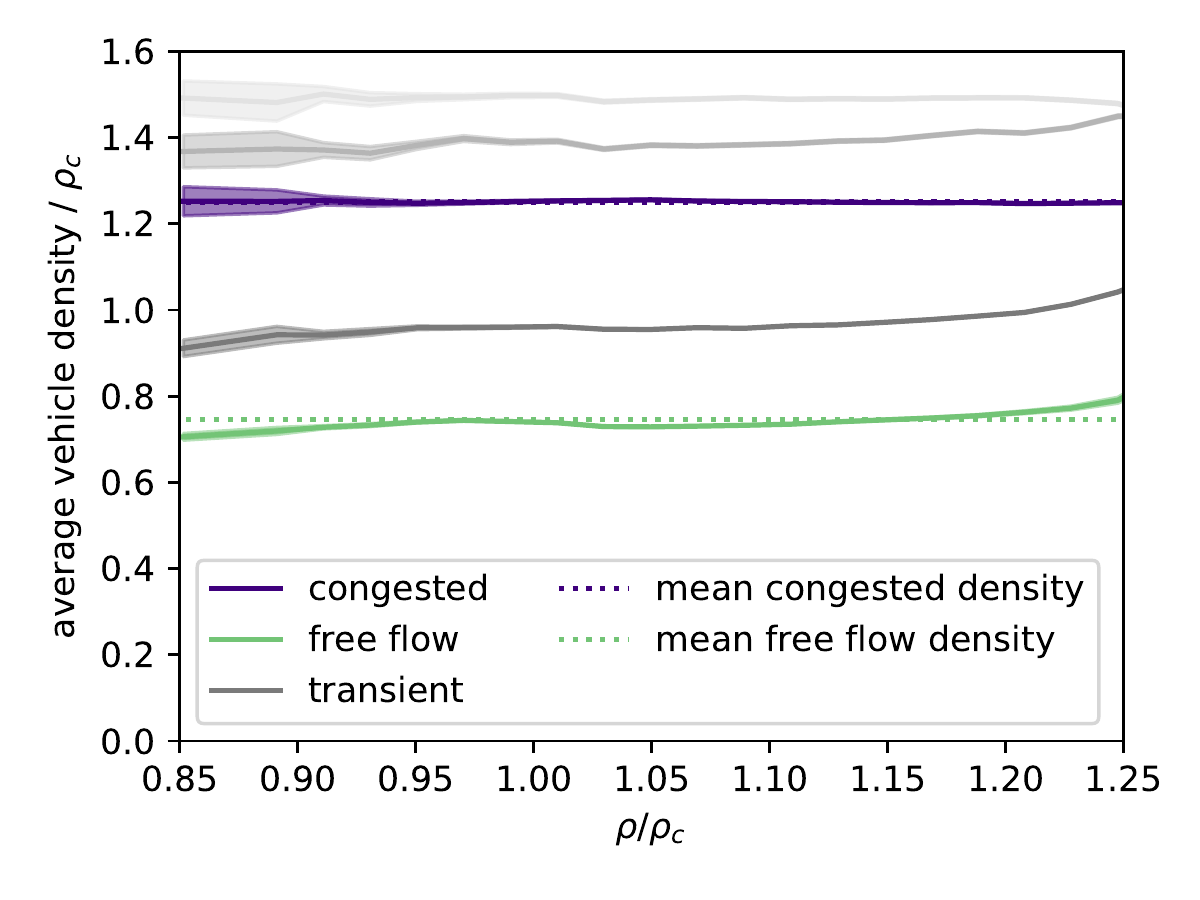}
    \caption{\textbf{Conditional vehicle density is independent from overall vehicle density.} A combination of the methods described in \Cref{fig:supp:classification} in \Cref{fig:congestion_probability} allows the calculation of the average vehicle density at charging stations that are congested (purple), in free flow (green) or at boundaries of the jam (grey). The conditional densities are approximately independent of the overall vehicle density $\rho$. We detect only small changes at large densities $\rho$, which are likely attributed to the finite sensitivity of the classification algorithms.}
    \label{fig:constant_average_density}
\end{figure}

We classify the state of a sample station for all times and compute average, variance and the number of datapoints conditional on each state. We combine the statistics for multiple simulations with the same total vehicle density $\rho$ to obtain the overall average of the vehicle density in each state, illustrated in \Cref{fig:constant_average_density}. For each state, the conditional vehicle density is approximately constant. However, at large overall vehicle densities, the free-flow density and two transient densities increase. This is likely a result of the imperfect clustering and classification method. At higher overall densities only a small fraction of the system is in free flow. When the regions are to small, the detection algorithm struggles detecting these regions.

\clearpage
\subsection*{Robustness}
In the main manuscript, we illustrated the emergence of congestion waves in EV charging infrastructure as a result of the queuing behavior of drivers. Here we demonstrate that this congested state robustly emerges for a wide range of parameters. In particular congestion waves emerge also for considerably lower charging thresholds than shown in Fig. 2 in the main manuscript. Additionally, congestion also occurs for heterogeneous vehicles or charging infrastructure, as long as the heterogeneity is sufficiently small, explained by an effective change of the fundamental diagram (see \Cref{fig:supp:fmd_heterogeneity} below).

\subsubsection*{Low charging thresholds on a highway}
\begin{figure}[h]
    \centering
    \includegraphics{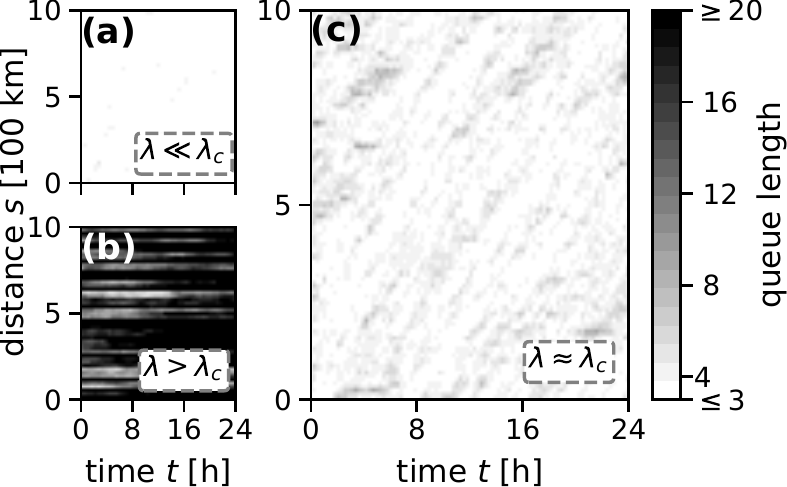}
    \caption{Queue lengths along a highway for a charging threshold $q^* = 0.1 q_{\max}$. All other parameters are identical to the ones used in Fig. 2 in the main manuscript. (a) For $\lambda = 150~\text{veh}/\text{h} \ll \lambda_c$ the system exhibits free flow. (b) For $\lambda = 300 ~\text{veh}/\text{h} > \lambda_{c}$ the system overloads and queues grow continuously. (c) Close to the critical rate $\lambda = 230 ~\text{veh}/\text{h} \approx \lambda_{c}$ congestion waves form.}
    \label{fig:q01}
\end{figure}

\begin{figure}[h]
    \centering
    \includegraphics{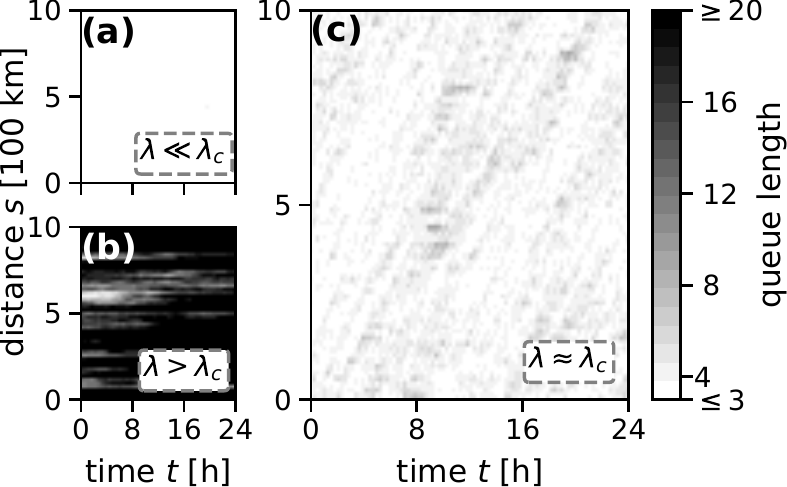}
    \caption{Queue lengths along a highway for a charging threshold $q^* = 0.2 q_{\max}$. All other parameters are identical to the ones used in Fig. 2 in the main manuscript. (a) For $\lambda = 150~\text{veh}/\text{h} \ll \lambda_c$ the system exhibits free flow. (b) For $\lambda = 300 ~\text{veh}/\text{h} > \lambda_c$ the system overloads and queues grow continuously. (c) Close to the critical rate $\lambda = 230 ~\text{veh}/\text{h} \approx \lambda_c$ congestion waves form.}
    \label{fig:q02}
\end{figure}

In Fig. 2 in the main manuscript we highlighted the congestion waves emerging in a highway like system, with vehicles entering and leaving the highway at random. For clarity of the argument, we set a intermediate charging threshold. Congestion waves also emerge for lower more realistic charging thresholds. In the following we show the congestion waves emerging close to the critical vehicle density also for charging thresholds $0.1q_{\max}$ (see Fig. \ref{fig:q01}) and $0.2q_{\max}$ (see Fig. \ref{fig:q02}).

\subsubsection*{Heterogeneous vehicles}
As an increasing number of car manufacturers produce electric vehicles, the pool of different models for electric vehicles is growing. The result is a diverse landscape of EV models with different charging speeds $\xi$ and ranges $q_\mathrm{max}$.
Additionally drivers may have different levels of confidence in their vehicles leading to different charging thresholds $q^*$. Here we simulate a few scenarios in which these parameters are distributed uniformly across an interval and show that congestion waves also occur under these heterogeneous conditions. For these simulations we consider roads with a fixed number of vehicles, matching the system described in Figures 3 and 4 in the main manuscript.

\begin{figure}[!h]
    \centering
    \includegraphics{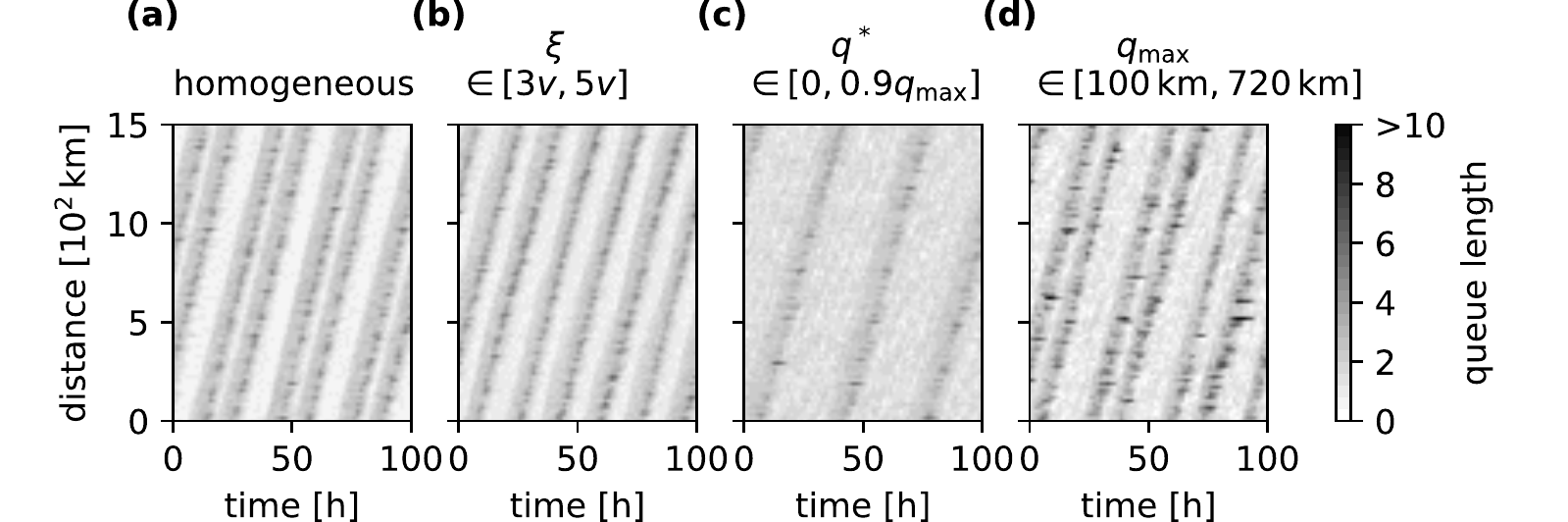}
    \caption{Congestion waves still emerge when vehicles or driver are comprised of heterogeneous populations. (a) A homogeneous system in which each vehicle is the same. (b) The charging threshold is distributed uniformly in an interval $[3v,5v]$. (c) The range of vehicles is distributed uniformly in the interval $[0,0.9q_\mathrm{max}]$. (d) The range of vehicles is distributed uniformly in the interval $[100\,\mathrm{km},720\,\mathrm{km}]$.}
    \label{fig:vehicle_heterogeneity}
\end{figure}

\subsubsection*{Heterogeneous infrastructure}
In a realistic setting, also the charging infrastructure will not be homogeneous. Charging stations will not be located at equal distances along a highway, and charging speed and the number of ports can vary. Here we simulate the system for different scenarios and show that congestion waves emerge.

We also show, that heterogeneities of certain magnitudes can introduce bottlenecks into the system which prevent the emergence of congestion waves by causing the system to overload at the weakest links. Such overloads are in themselves undesirable outcomes, thus charging stations added in the future will likely resolve these bottlenecks and charging station distribution will become more homogeneous.\\

\paragraph{Heterogeneous charging ports}

We introduce heterogeneities in the number of charging ports at stations by picking each station at random from two types of stations, one with $4-\Delta p$ charging ports and one with $4+\Delta p$ charging ports, where $\Delta p \in \{0,1,2,3\}$. The average number of charging ports therefore stays constant, as does the critical density.

\begin{figure}[h]
    \centering
    \includegraphics{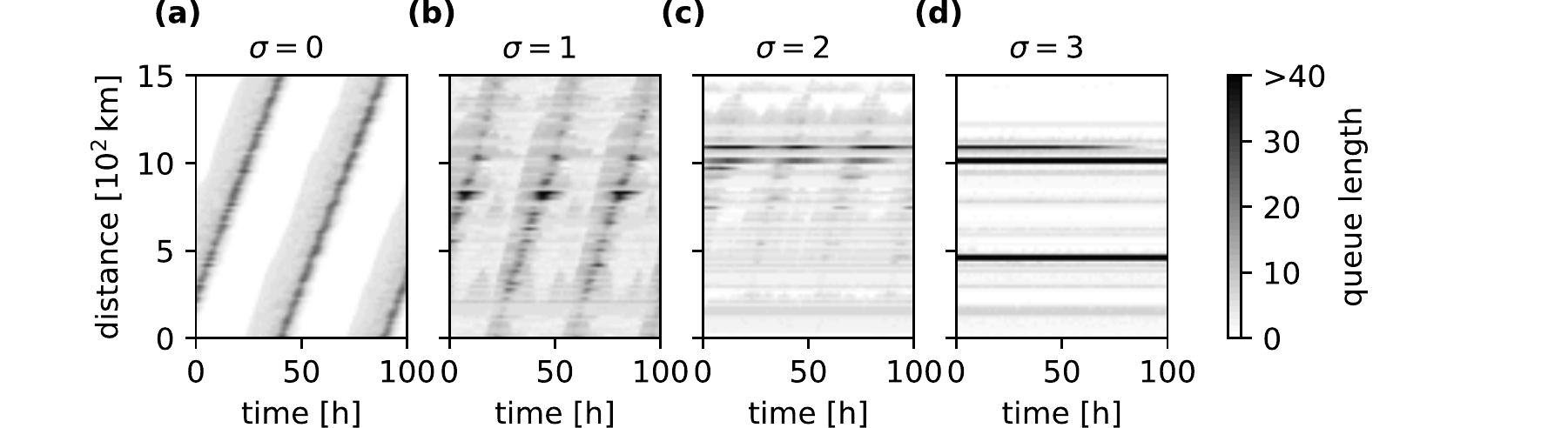}
    \caption{Heterogeneous port distribution. Stations are divided into two types equipped with different port numbers. (a) Every charging station is equipped with 4 ports. (b) Half of the stations are equipped with 3 ports and the other with 5 ports. (c) Stations are equipped with 2 or 6 ports. (d) Stations are equipped with 1 or 7 ports. As the system becomes more heterogeneous the congestion patterns become distorted until the queues of single overloaded charging stations dominate the flow.}
    \label{fig:supp:port_heterogeneity}
\end{figure}

\Cref{fig:supp:port_heterogeneity} shows that increasing the variance of the ports distorts the congestion waves. When the heterogeneities are too large some stations become permanently overloaded. Overloaded stations bind a large number of vehicles in their queue, which reduces the effective density everywhere outside this station and prevents congestion waves.\\

\paragraph{Heterogeneous distances}

Next, we introduce heterogeneities in the distance between stations, which is likely to be the case in densely populated regions along a highway, because of the presence of cities and coinciding infrastructure.

Like in the previous case we model heterogeneous road conditions by choosing the distances between stations from two different possibilities, $15\,\mathrm{km}-\Delta l$ and $15\,\mathrm{km}+\Delta l$. We choose the values of $\Delta l$ such that the relative standard deviation
\begin{align*}
     \frac{\sigma}{15\,\mathrm{km}} = \frac{\Delta l}{15\,\mathrm{km}} \in \{ 0,\frac{1}{4},\frac12,\frac34 \}.
\end{align*}
This corresponds to the same as for the heterogeneous port distribution.

\begin{figure}[h]
    \centering
    \includegraphics{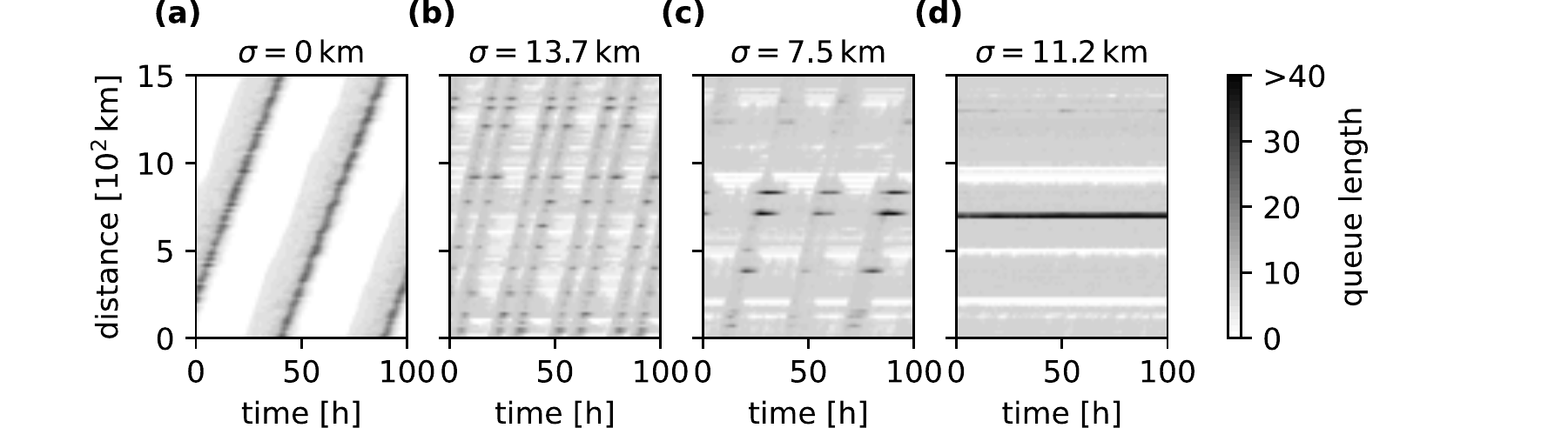}
    \caption{Heterogeneous distances between stations (a) All distances between charging stations are 15\,km. (b)-(d) Edges between stations are either 11.25\,km or 18.75\,km (b), 7.5\,km or 22.5\,km (c), or 3.75\,km and 26.25\,km (d) with 50\% chance. The distances were chosen to match the relative standard deviation in \Cref{fig:supp:port_heterogeneity}. Like in \Cref{fig:supp:port_heterogeneity}, increasing heterogeneity distorts the congestion patterns and leads to the emergence of singular overloaded stations at large standard deviations.
    }
    \label{fig:supp:distance_heterogeneity}
\end{figure}

The results depicted in \Cref{fig:supp:distance_heterogeneity} are comparable to the case in which the number of charging ports is distributed heterogeneously. Heterogeneities distort the congestion patterns, and when the deviations are large enough bottlenecks emerge binding a large portion of vehicles to a single station.

\begin{figure}
    \centering
    \includegraphics{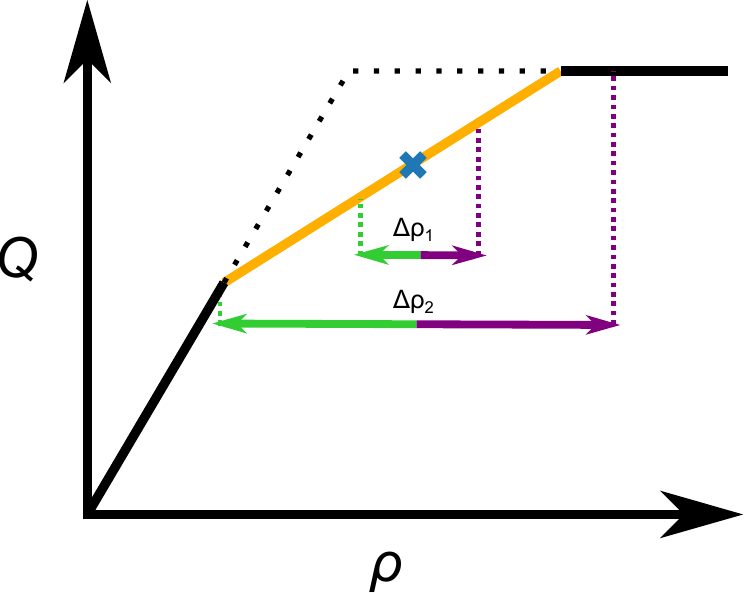}
    \caption{Effects of perturbations on EV traffic flow. Fluctuations in the number of ports, or the distance between charging stations changes the local vehicle density per station or length. In the phase space this corresponds to a shift in flow according to the fundamental diagram. Small fluctuations ($\Delta \rho_1$) remain on the congested branch of the fundamental diagram while large enough fluctuations ($\Delta \rho_2$) either produce free-flow regions, i.e. regions where congestion dissipates (see white stripes in \Cref{fig:supp:port_heterogeneity,fig:supp:distance_heterogeneity} (c) and (d)), or overloaded regions, where some charging stations drastically overload (see black lines in \Cref{fig:supp:port_heterogeneity,fig:supp:distance_heterogeneity} (c) and (d)).
    }
    \label{fig:supp:fmd_heterogeneity}
\end{figure}

The effects of perturbations can be explained by the application of the fundamental diagram in \Cref{fig:supp:fmd_heterogeneity}. Without perturbation a system that exhibits congestion waves rests on the congestion branch (cross, orange line). When the number of ports is locally increased the ratio of vehicles to ports is reduced. When the number of ports is decreased the ratio of vehicles to ports is increased. As long as the perturbed state is still on the congested branch, which is the case for small heterogeneities, congestion waves emerge. When the perturbation becomes too large, the system can either become locally overloaded, which creates bottlenecks, or the system can become locally free, which leads to the dissipation of congestion that might have originated upstream.

\newpage

\end{document}